\documentclass[aps,pra,longbibliography,superscriptaddress,twocolumn]{revtex4-1} 

\usepackage[T1]{fontenc}
\usepackage{graphicx,amsmath,amssymb,amsfonts,dsfont}
\usepackage[pdftex]{hyperref}
\usepackage{physics}
\usepackage{bm}
\usepackage[framemethod=TikZ]{mdframed}

\newenvironment{Frame}[1][]{%
    \begin{mdframed}[%
        frametitle={#1},
        skipabove=\baselineskip plus 2pt minus 1pt,
        skipbelow=\baselineskip plus 2pt minus 1pt,
        linewidth=0.5pt,
        frametitlerule=true,
        frametitlebackgroundcolor=gray!30,
        roundcorner=10pt,
        nobreak=true
    ]%
}{%
    \end{mdframed}
}

\def\Tr{\mathrm{Tr}}
\def\ket#1{|#1\rangle}

\def\bra#1{\langle#1|}

\def\ah{\hat{a}}
\def\ad{\hat{a}^\dagger}
\def\xh{\hat{x}}
\def\ph{\hat{p}}
\def\qh{\hat{q}}

\def\rh{\hat{\bm{r}}}
\def\rb{\bm{r}}
\def\rd{\bar{\bm{r}}}
\def\bs{\bm{\sigma}}

\def\Fb{\bm{F}}
\def\db{\bm{d}}

\begin{document}

\title{Gaussian states and operations -- a quick reference}

\author{Jonatan Bohr Brask}\email{jonatan.brask@fysik.dtu.dk}\affiliation{Department of Physics, Technical University of Denmark, Fysikvej, Kongens Lyngby 2800, Denmark}

\date{\today}

\begin{abstract}
Gaussian quantum states of bosonic systems are an important class of states. In particular, they play a key role in quantum optics as all processes generated by Hamiltonians up to second order in the field operators (i.e.~linear optics and quadrature squeezing) preserve Gaussianity. A powerful approach to calculations and analysis of Gaussian states is using phase-space variables and symplectic transformations. The purpose of this note is to serve as a concise reference for performing phase-space calculations on Gaussian states. In particular, we list symplectic transformations for commonly used optical operations (displacements, beam splitters, squeezing), and formulae for tracing out modes, treating homodyne measurements, and computing fidelities.
\end{abstract}

\maketitle

\section{Sources}

There are many excellent sources on the analysis of Gaussian states and transformation and their use in quantum optics and quantum information science, which provide much more thorough treatments of the subject than presented here. The present note should be thought of only as a quick-reference collection of fomulae to aid in calculations. As such, I will not generally provide citations to the original work deriving each of the results presented. For thorough references to primary sources and historical overview, readers may refer to the reviews \cite{Braunstein2005,Weedbrook2012}. In addition to these reviews, I draw on the excellent lecture notes of Ferraro, Olivares, and Paris \cite{Ferraro2005}, as well as the PhD thesis of Zhang \cite{Zhang2018}, the books by Gerry and Knight \cite{Gerry2005} and Serafini \cite{Serafini2017}, and on the work of Banchi, Braunstein, and Pirandola on computing fidelity for arbitrary Gaussian states \cite{Banchi2015}.

\section{Basic concepts and notation}

\subsubsection{Mode and quadrature operators}

We will consider bosonic systems consisting of $n$ harmonic oscillators, with corresponding annihilation and creation operators $\ah_k$, $\ad_k$ with $k=1,\ldots,n$. We also refer to $\ah_k$ as \textit{mode operators}. Throughout the note, we set $\hbar = 1$ and adopt the canonical commutation relations
\begin{align}
[\ah_k,\ah_l] & = [\ad_k,\ad_l] = 0 , \\
[\ah_k,\ad_l] & = \delta_{kl} . 
\end{align}
We define Hermitian position- and momentum-like operators for each mode
\begin{equation}
\label{eq.xp}
\xh_k = \frac{1}{\sqrt{2}}(\ah_k + \ad_k) , \hspace{1cm} \ph_k = \frac{1}{i\sqrt{2}}(\ah_k - \ad_k) ,
\end{equation}
with the canonical commutator
\begin{equation}
[\xh_k,\ph_l] = i \delta_{kl} .
\end{equation}
Note that different conventions for the normalisation in Eq.~\eqref{eq.xp} can be found in the literature (for instance, Ref.~\cite{Weedbrook2012} uses 1 rather than $1/\sqrt{2}$, and Ref.~\cite{Gerry2005} uses $1/2$). We will stick to the canoncial convention above throughout. Following the convention in quantum optics, we also refer to $\xh_k$ and $\ph_k$ as \textit{quadrature operators}, and we define the general, rotated quadrature
\begin{align}
\qh_k(\phi) & = \frac{1}{\sqrt{2}}(\ah_k e^{-i\phi} + \ad_k e^{i\phi}) \\
& = \cos(\phi)\xh_k + \sin(\phi)\ph_k ,
\end{align}
such that $\qh_k(0) = \xh_k$ and $\qh_k(\frac{\pi}{2})=\ph_k$. For the commutator, we have
\begin{equation}
[\qh_k(\phi),\qh_l(\varphi)] = i \delta_{kl} \sin(\varphi - \phi) .
\end{equation}
It will be convenient to collect the $2n$ quadrature operators for the $n$ modes into a vector. We define the column vector
\begin{equation}
\rh = \begin{pmatrix}
\xh_1 \\ \ph_1 \\ \vdots \\ \xh_n \\ \ph_n 
\end{pmatrix} .
\end{equation}
The canonical commutation relations can then be written
\begin{equation}
[\rh_k,\rh_l] = i \bm{\Omega}_{kl} ,
\end{equation}
where $\bm{\Omega}$ is the symplectic matrix defined by
\begin{equation}
\bm{\Omega} = \bigoplus_{k=1}^{2n} \begin{pmatrix}
0 & 1 \\ -1 & 0
\end{pmatrix} = \openone_n \otimes \begin{pmatrix}
0 & 1 \\ -1 & 0
\end{pmatrix} ,
\end{equation}
where $\openone_n$ is the $n\times n$ identity matrix.

\subsubsection{Displacement operators}

The displacement operator on a single mode $k$ is
\begin{equation}
\hat{D}_k(\alpha) = e^{\alpha \ad_k - \alpha^* \ah_k} ,
\end{equation}
where $\alpha$ is a complex number. It transforms the mode operator as
\begin{equation}
\label{eq.disp}
\hat{D}_k^\dagger(\alpha) \ah_k \hat{D}_k(\alpha) = \ah_k + \alpha .
\end{equation}
When applied to the vacuum state (ground state), it generates the coherent states $\ket{\alpha} = \hat{D}(\alpha)\ket{vac}$.

The joint displacement operator on $n$ modes is
\begin{equation}
\hat{D}(\bm{\alpha}) = \hat{D}_1(\alpha_1)\otimes\cdots\otimes\hat{D}_n(\alpha_n) ,
\end{equation}
where $\bm{\alpha} = (\alpha_1,\ldots,\alpha_n)$. Using the vector of quadrature operators and the symplectic matrix above, we can equivalently express the joint displacement operator as a function of $2n$ real variables, collected in a vector $\bm{r} = (x_1,p_1,\ldots,x_n,p_n)^T$, as
\begin{equation}
\hat{D}(\bm{r}) = \exp(i \rh^T \bm{\Omega} \bm{r}) ,
\end{equation}
with the relation $\alpha_k = \frac{x_k + i p_k}{\sqrt{2}}$ to the complex displacements.

\subsubsection{Covariance matrices}

Given a quantum state $\rho$ of an $n$-mode system, the corresponding covariance matrix $\bs$ has elements
\begin{Frame}[Covariance matrix]
\begin{equation}
\sigma_{kl} = \frac{1}{2}\langle \{\rh_k,\rh_l\}\rangle -  \langle\rh_k\rangle\langle\rh_l\rangle 
\end{equation}
\end{Frame}
where $\{\cdot,\cdot\}$ denotes the anticommutator and $\langle \hat{A} \rangle = \Tr[\hat{A} \rho]$ is the expectation value.

The covariance matrix (for valid density matrices $\rho$) is real, symmetric, positive definite $\bs>0$, and fulfills
\begin{equation}
\bs + \frac{i}{2}\bm{\Omega} \geq 0 .
\end{equation}

Note that the diagonal elements of $\bs$ are the variances of the quadrature operators, while non-zero off-diagonal elements correspond to correlations between quadratures.

\subsubsection{Wigner functions and characteristic functions}

We let $\bm{x}$ and $\bm{p}$ denote $n$-dimensional real vectors. Furthermore, let $\ket{x_k}$ denote an eigenstate of $\xh_k$ with eigenvalue $x_k$, and $\ket{\bm{x}} = \ket{x_1,x_2,\ldots,x_n} = \ket{x_1}\ket{x_2}\cdots\ket{x_n}$. The Wigner function corresponding to state $\rho$ is then defined by
\begin{equation}
W(\bm{x},\bm{p}) = \frac{1}{(2\pi)^n} \int_{\mathbb{R}^n} \bra{\bm{x} + \frac{1}{2}\bm{q}}\rho\ket{\bm{x} - \frac{1}{2}\bm{q}} e^{i \bm{p}\cdot\bm{q}} \, d\bm{q} .
\end{equation}
We may also write $W(\bm{r})$ for $\bm{r} = (x_1,p_1,\ldots,x_n,p_n)$, where it is understood that $W(\bm{r}) = W(\bm{x},\bm{p})$. Here, the $x_k$ and $p_k$ are phase-space variables, and the Wigner function is a phase-space representation of the state $\rho$. All the information contained in $\rho$ is also contained in $W$ and vice versa, i.e.~the Wigner function is an equivalent representation of the quantum state of the system.

Another equivalent representation of $\rho$ is the characteristic function of $n$ complex variables $\alpha_1,\ldots,\alpha_n$
\begin{equation}
\chi(\bm{\alpha}) = \Tr[\rho \hat{D}(\bm{\alpha})] ,
\end{equation}
or equivalently of $2n$ real variables $\chi(\bm{r}) = \Tr[\rho \hat{D}(\bm{r})]$. The Wigner function can also be expressed as the Fourier transform of the characteristic function
\begin{equation}
W(\bm{r}) = \frac{1}{(2\pi)^{2n}} \int_{\mathbb{R}^{2n}} \exp(-i\bm{r}^T \bm{\Omega}  \bm{s}) \chi(\bm{s}) d\bm{s} .
\end{equation}

\subsubsection{Gaussian states}

Gaussian states are those states for which the Wigner function is a (multivariate) Gaussian function.

Gaussian states are completely determined by the first and second moments of the quadrature operators. That is, by the vector of expectation values $\rd = \langle \rh \rangle$ and the covariance matrix $\bs$. We refer to $\rd$ as the \textit{displacement vector} of the state. In terms of the displacement vector and covariance matrix, the Wigner function of a Gaussian state can be expressed
\begin{Frame}[Wigner function (Gaussian state)]
\begin{equation}
\label{eq.Wfctgauss}
W(\rb) = \frac{1}{(2\pi)^n\sqrt{\det(\bs)}} e^{-\frac{1}{2} (\rb - \rd)^T \bs^{-1} (\rb - \rd)} 
\end{equation} 
\end{Frame}
where $\det(\bs)$ and $\bs^{-1}$ denote the determinant and inverse of $\bs$, respectively. Note that the Wigner function is normalised
\begin{equation}
\int_{\mathbb{R}^{2n}} W(\rb) d\rb = 1.
\end{equation}

\subsubsection{Gaussian operations}

Gaussian operations are those which take Gaussian states to Gaussian states, i.e.~those which preserve Gaussianity.

Since Gaussian states are completely determined by $\rd$ and $\bs$, Gaussian operations are then determined by how they transform $\rd$ and $\bs$. Unitary Gaussian operations correspond to those which can be generated by Hamiltonians which are (at most) quadratic in the mode operators $\ah_k$ and $\ad_k$. These correspond exactly to symplectic transformations of the displacement vector and covariance matrix. They map
\begin{equation}
\bs \rightarrow \Fb \bs \Fb^T , \hspace{0.5cm} \text{and} \hspace{0.5cm} \rd \rightarrow \Fb \rd + \db ,
\end{equation}
where $\db$ is a $2n$-dimensional real vector of displacements, and the matrix $\Fb$ fulfills
\begin{equation}
\Fb\bm{\Omega}\Fb^T = \bm{\Omega} .
\end{equation}
Gaussian unitaries are thus determined by a symplectic matrix $\Fb$ and a displacement vector $\db$. Every Gaussian unitary corresponds to a symplectic transformation and, conversely, for every symplectic transformation there exist a quadratic Hamiltonian and corresponding Gaussian unitary which generates it.

It follows from the Euler decomposition that every Gaussian unitary on $n$ modes can be realised by a passive, linear transformation followed by single-mode squeezing in each mode, followed by another passive, linear transformation. That is, in the language of quantum optics, by combining beam splitters, phase shifts, and single-mode squeezing.

\section{Gaussian unitaries}

In this section, we list symplectic transformations corresponding to some commonly encountered Gaussian unitary operations.

\subsubsection{Displacements}

From \eqref{eq.disp}, a displacement on a single mode $k$ transforms the quadrature operators as
\begin{align}
\hat{D}^\dagger(\alpha) \xh \hat{D}(\alpha) & = \xh + \frac{\alpha + \alpha^*}{\sqrt{2}} = \xh_k + \sqrt{2}\Re(\alpha) , \\
\hat{D}^\dagger(\alpha) \ph \hat{D}(\alpha) & = \ph + \frac{\alpha - \alpha^*}{i\sqrt{2}} = \xh_k + \sqrt{2}\Im(\alpha) .
\end{align}
For an $n$-mode state and complex displacements $\bm{\alpha} = (\alpha_1,\ldots,\alpha_n)$, it follows that
\begin{align}
\bs & \rightarrow \bs \hspace{0.5cm} \text{(the covariance matrix is unchanged)} , \\
\rd & \rightarrow \rd + \sqrt{2}(\Re(\alpha_1) , \Im(\alpha_1) , \ldots , \Re(\alpha_n) , \Im(\alpha_n)^T .
\end{align}
As a symplectic transformation therefore, a displacement defined by an $n$-dimensional complex vector $\bm{\alpha}$ corresponds to
\begin{Frame}[Displacement]
\begin{align}
\Fb & = \openone_{2n}  , \\
\db & = \sqrt{2}\begin{pmatrix}
\Re(\alpha_1) \\ \Im(\alpha_1) \\ \vdots \\ \Re(\alpha_n) \\ \Im(\alpha_n)
\end{pmatrix} .
\end{align}
\end{Frame}

\subsubsection{Phase shifts}

A phase shift on a single mode by an angle $\phi$ transforms the mode operators as
\begin{equation}
\ah \rightarrow e^{-i\phi} \ah , \hspace{1cm} \ad \rightarrow e^{i\phi} \ad .
\end{equation}
The quadrature operators hence transform as
\begin{align}
\xh & \rightarrow \cos(\phi)\xh + \sin(\phi)\ph = \xh(\phi), \\
\ph & \rightarrow \cos(\phi)\ph - \sin(\phi)\xh = \xh(\phi + \frac{\pi}{2}).
\end{align}
As a symplectic transformation on a single mode therefore, the phase shift corresponds to a rotation
\begin{Frame}[Phase shift]
\begin{align}
\Fb & = \begin{pmatrix}
\cos(\phi) & \sin(\phi) \\
-\sin(\phi) & \cos(\phi) 
\end{pmatrix} \equiv \bm{R}(\phi) , \\
\db & = \bm{0} .
\end{align}
\end{Frame}

When phase shifts $\phi_1,\ldots,\phi_n$ are applied to each mode of an $n$-mode system, we have
\begin{align}
\Fb & = \bigoplus_{k=1}^n \bm{R}(\phi_k) , \\
\db & = \bm{0} .
\end{align}

\subsubsection{Beam splitters}

We consider a beam splitter with transmittivity $\eta$, which transforms the mode operators $\ah_1$, $\ah_2$ of two modes as
\begin{align}
\ah_1 & \rightarrow  \sqrt{\eta}\,\ah_1 + \sqrt{1-\eta}\,\ah_2 , \\
\ah_2 & \rightarrow -\sqrt{1-\eta}\,\ah_1 + \sqrt{\eta}\,\ah_2 .
\end{align}
That is
\begin{equation}
\begin{pmatrix}
\ah_1 \\ \ah_2 
\end{pmatrix} \rightarrow \begin{pmatrix}
\sqrt{\eta} & \sqrt{1-\eta} \\
-\sqrt{1-\eta} & \sqrt{\eta}
\end{pmatrix} \begin{pmatrix}
\ah_1 \\ \ah_2 
\end{pmatrix} .
\end{equation}
The quadrature operators then transform as
\begin{equation}
\begin{pmatrix}
\xh_1 \\ \ph_1 \\ \xh_2 \\ \ph_2
\end{pmatrix} \rightarrow \begin{pmatrix}
\sqrt{\eta} & 0 & \sqrt{1-\eta} & 0 \\
0 & \sqrt{\eta} & 0 & \sqrt{1-\eta} \\
-\sqrt{1-\eta} & 0 & \sqrt{\eta} & 0 \\
0 & -\sqrt{1-\eta} & 0 & \sqrt{\eta} 
\end{pmatrix} \begin{pmatrix}
\xh_1 \\ \ph_1 \\ \xh_2 \\ \ph_2
\end{pmatrix} .
\end{equation}
The beam splitter thus does not contribute any displacement but it does mix the mode operators. As a symplectic transformation on two modes, we have
\begin{Frame}[Beam splitter]
\begin{align}
\Fb & = \left( \begin{array}{rr} \sqrt{\eta} \openone_2 & \sqrt{1-\eta} \openone_2 \\ -\sqrt{1-\eta} \openone_2 & \sqrt{\eta} \openone_2 \end{array} \right) \equiv \Fb_{\eta} , \\
\db & = \bm{0} .
\end{align}
\end{Frame}

When acting on a pair of modes (say $k,l$) of an $n$-mode system, the symplectic matrix should be equal to $\Fb_{\eta}$ in the $kl$-subspace and identity on the rest of the space. Let $\bm{P}$ be a permutation matrix shifting modes $k$ and $l$ to the first two modes, i.e.\ $P(x_1,p_1,\ldots,x_n,p_n)^T = (x_k,p_k,x_l,p_l,\ldots)^T$. Then
\begin{align}
\Fb & = \bm{P}^{-1} \begin{pmatrix}
\Fb_{\eta} & 0 \\
0 & \openone_{2(n-2)}
\end{pmatrix} \bm{P} , \\
\db & = \bm{0} .
\end{align}

\subsubsection{Single-mode squeezing}

The single-mode squeezing operator is
\begin{equation}
\hat{S}(\xi) = e^{\frac{1}{2}(\xi^* \ah^2 - \xi (\ad)^2)} ,
\end{equation}
with $\xi = r e^{i \theta}$. The mode operators then transform as
\begin{align}
\hat{S}(\xi)^\dagger \ah \hat{S}(\xi) & = \cosh(r) \ah - e^{i\theta} \sinh(r) \ad , \\
\hat{S}(\xi)^\dagger \ad \hat{S}(\xi) & = \cosh(r) \ad - e^{-i\theta} \sinh(r) \ah .
\end{align}
Hence, setting $c_r = \cosh(r)$, $s_r = \sinh(r)$, the quadrature operators transform as
\begin{align}
\hat{S}(\xi)^\dagger \xh \hat{S}(\xi) & = [c_r - \cos(\theta)s_r] \xh - \sin(\theta)s_r \ph , \\
\hat{S}(\xi)^\dagger \ph \hat{S}(\xi) & = [c_r + \cos(\theta)s_r] \ph - \sin(\theta)s_r \xh .
\end{align}
Note that this squeezes $\xh(\frac{\theta}{2})$ and anti-squeezes $\xh(\frac{\theta}{2} + \frac{\pi}{2})$. For example, for $\theta=0$ we get $\xh \rightarrow e^{-r} \xh$ and $\ph \rightarrow e^r \ph$.

As a symplectic transformation on a single mode, we have
\begin{Frame}[Single-mode squeezing]
\begin{align}
\Fb & = \cosh(r) \openone_2 - \sinh(r) \bm{S}_\theta \equiv \Fb_{\xi} , \\
\db & = \bm{0} ,
\end{align}
where
\begin{equation}
\bm{S}_\theta = \begin{pmatrix}
\cos(\theta) & \sin(\theta) \\
\sin(\theta) & -\cos(\theta)
\end{pmatrix} .
\end{equation}
\end{Frame}

When single-mode squeezing with parameters $\xi_1,\ldots,\xi_n$ are applied to each mode of an $n$-mode system, we have
\begin{align}
\Fb & = \bigoplus_{k=1}^n \Fb_{\xi_k} , \\
\db & = \bm{0} .
\end{align}

\subsubsection{Two-mode squeezing}

The two-mode squeezing operator acting on modes 1,2 is
\begin{equation}
\hat{S}_2(\xi) = e^{\xi^* \ah_1 \ah_2 - \xi \ad_1 \ad_2} ,
\end{equation}
with $\xi = r e^{i \theta}$. It transforms the mode operators according to
\begin{align}
\hat{S}_2(\xi)^\dagger \ah_1 \hat{S}_2(\xi) & = \cosh(r) \ah_1 - e^{i\theta} \sinh(r) \ad_2 , \\
\hat{S}_2(\xi)^\dagger \ad_1 \hat{S}_2(\xi) & = \cosh(r) \ad_1 - e^{-i\theta} \sinh(r) \ah_2 , \\
\hat{S}_2(\xi)^\dagger \ah_2 \hat{S}_2(\xi) & = \cosh(r) \ah_2 - e^{i\theta} \sinh(r) \ad_1 , \\
\hat{S}_2(\xi)^\dagger \ad_2 \hat{S}_2(\xi) & = \cosh(r) \ad_2 - e^{-i\theta} \sinh(r) \ah_1 .
\end{align}
Setting $c_r = \cosh(r)$, $s_r = \sinh(r)$, the quadratures then transform as
\begin{align}
\hat{S}_2(\xi)^\dagger \xh_1 \hat{S}_2(\xi) & = c_r \xh_1 - \cos(\theta)s_r \xh_2 -\sin(\theta) s_r \ph_2 , \\
\hat{S}_2(\xi)^\dagger \ph_1 \hat{S}_2(\xi) & = c_r \ph_1 + \cos(\theta)s_r \ph_2 -\sin(\theta) s_r \xh_2 , \\
\hat{S}_2(\xi)^\dagger \xh_2 \hat{S}_2(\xi) & = c_r \xh_2 - \cos(\theta)s_r \xh_1 -\sin(\theta) s_r \ph_1 , \\
\hat{S}_2(\xi)^\dagger \ph_2 \hat{S}_2(\xi) & = c_r \ph_2 + \cos(\theta)s_r \ph_1 -\sin(\theta) s_r \xh_1 .
\end{align}
Note that for $\theta=0$, this squeezes both $\xh_1+\xh_2$ and $\ph_1 - \ph_2$ since $\xh_1+\xh_2 \rightarrow e^{-r}(\xh_1+\xh_2)$ and $\ph_1-\ph_2 \rightarrow e^{-r}(\ph_1+\ph_2)$.

As a symplectic transformation on two modes, we have
\begin{Frame}[Two-mode squeezing]
\begin{align}
\Fb & = \begin{pmatrix}
\cosh(r) \openone_2 & -\sinh(r) \bm{S}_\theta \\
-\sinh(r) \bm{S}_\theta &  \cosh(r) \openone_2
\end{pmatrix} \equiv \Fb_{2,\xi} \\
\db & = \bm{0} , 
\end{align}
where again
\begin{equation}
\bm{S}_\theta = \begin{pmatrix}
\cos(\theta) & \sin(\theta) \\
\sin(\theta) & -\cos(\theta)
\end{pmatrix} .
\end{equation}
\end{Frame}

When acting on a pair of modes (say $k,l$) of an $n$-mode system, the symplectic matrix should be equal to $\Fb_{2,\xi}$ in the $kl$-subspace and identity on the rest of the space. Let $\bm{P}$ be a permutation matrix shifting modes $k$ and $l$ to the first two modes, i.e.\ $P(x_1,p_1,\ldots,x_n,p_n)^T = (x_k,p_k,x_l,p_l,\ldots)^T$. Then
\begin{align}
\Fb & = \bm{P}^{-1} \begin{pmatrix}
\Fb_{2,\xi} & 0 \\
0 & \openone_{2(n-2)}
\end{pmatrix} \bm{P} , \\
\db & = \bm{0} .
\end{align}

\subsubsection{Controlled phase gate}

A controlled phase gate acting on modes 1,2 is defined by the operator $CZ(\phi) = e^{i \phi x_1 x_2}$. The quadratures transform according to
\begin{align}
CZ(\phi)^\dagger \xh_1 CZ(\phi) & = \xh_1 , \\
CZ(\phi)^\dagger \ph_1 CZ(\phi) & = \ph_1 + \phi \xh_2 , \\
CZ(\phi)^\dagger \xh_2 CZ(\phi) & = \xh_2 , \\
CZ(\phi)^\dagger \ph_2 CZ(\phi) & = \ph_2 + \phi \xh_1 , \\ .
\end{align}
As a symplectic transformation on a pair of modes therefore
\begin{Frame}[Controlled phase]
\begin{align}
\Fb & = \begin{pmatrix}
1 & 0 & 0 & 0 \\
0 & 1 & \phi & 0 \\
0 & 0 & 1 & 0 \\
\phi & 0 & 0 & 1
\end{pmatrix} \equiv \Fb_{\phi} \\
\db & = \bm{0} , 
\end{align}
\end{Frame}

When acting on a pair of modes (say $k,l$) of an $n$-mode system, the symplectic matrix should be equal to $\Fb_{\phi}$ in the $kl$-subspace and identity on the rest of the space. Let $\bm{P}$ be a permutation matrix shifting modes $k$ and $l$ to the first two modes, i.e.\ $P(x_1,p_1,\ldots,x_n,p_n)^T = (x_k,p_k,x_l,p_l,\ldots)^T$. Then
\begin{align}
\Fb & = \bm{P}^{-1} \begin{pmatrix}
\Fb_{\phi} & 0 \\
0 & \openone_{2(n-2)}
\end{pmatrix} \bm{P} , \\
\db & = \bm{0} .
\end{align}

\section{Other Gaussian operations}

\subsubsection{Tracing out}

Consider a bidivision of an $n$-mode system into subsystems A and B and order the modes such that $\rh = (\rh_A , \rh_b )^T$, where $\rh_A$ and $\rh_B$ are vectors of quadratures for A and B, respectively. 

For any state $\rho$ of the joint system and any operator $\hat{O}_A$ acting only in A, we have
\begin{equation}
\langle \hat{O}_A \rangle_\rho = \Tr[(\hat{O}_A\otimes\openone_B) \rho ] =\Tr_A[ \hat{O}_A \rho_A ] ,
\end{equation}
where $\rho_A$ is the reduced state of A. Similarly for B. It follows that the covariance matrix of the joint system can be written
\begin{equation}
\bs = \begin{pmatrix}
\bm{A} & \bm{C} \\
\bm{C}^T & \bm{B}
\end{pmatrix} ,
\end{equation}
where $\bm{A}$ and $\bm{B}$ are the covariance matrices corresponding to $\rho_A$ and $\rho_B$, respectively, and $\bm{C}$ encodes correlations between A and B. The displacement vector of the joint system is 
\begin{equation}
\rd = \langle \rh \rangle_\rho = \begin{pmatrix}
\langle \rh_A \rangle_{\rho_A} \\ 
\langle \rh_B \rangle_{\rho_B} 
\end{pmatrix}  = \begin{pmatrix}
\rd_A \\
\rd_B
\end{pmatrix} .
\end{equation}

When the state of the joint system is Gaussian, the reduced states $\rho_A$ and $\rho_B$ are also Gaussian and are simply defined by $\bs_A = \bm{A}$ and $\rd_A$ and by $\bs_B = \bm{B}$ and $\rd_B$, respectively.

\begin{Frame}[Tracing out]
For a joint system AB in a Gaussian state $\rho_{AB}$ with displacement vector $\rd = (\rd_A , \rd_b )^T$ and covariance matrix
\begin{equation}
\bs = \begin{pmatrix}
\bm{A} & \bm{C} \\
\bm{C}^T & \bm{B}
\end{pmatrix} ,
\end{equation}
the reduced states $\rho_A$, $\rho_B$ of subsystems A and B, respectively, are also Gaussian with displacement vectors $\rd_A$ and $\rd_B$ and covariance matrices $\bm{A}$ and $\bm{B}$.
\end{Frame}

Starting from an arbitrary ordering of modes, $\bs$ and $\rd$ can always be brought on the form above by applying an appropriate permutation reordering the modes with those of A first followed by those of B.

\subsubsection{Homodyne measurements}

Ideal homodyne measurements correspond to projective measurements of quadrature operators.\\

\noindent \textit{Measurements of $\xh$}\\

\noindent We consider an ideal homodyne measurement of $\xh_n$ on the last mode of an $n$-mode system in a Gaussian state. Denoting the measured mode subsystem B and the first $n-1$ modes A, we can write the covariance matrix and displacement vector of the state
\begin{align}
\bs = \begin{pmatrix}
\bm{A} & \bm{C} \\
\bm{C}^T & \bm{B}
\end{pmatrix} , \hspace{1cm} \rd = \begin{pmatrix}
\bm{a} \\
\bm{b}
\end{pmatrix} .
\end{align}
We would like to express the conditional state of A upon obtaining a particular measurement outcome as well as the probability distribution over outcomes in terms of elements of $\bs$ and $\rd$.

The covariance matrix of A after the measurement is \textit{independent} of the measurement outcome and is given by
\begin{equation}
\bs_A = \bm{A} - \bm{C} (\bm{\Pi}\bm{B}\bm{\Pi})^{-1} \bm{C}^T ,
\end{equation}
where
\begin{equation}
\bm{\Pi} = \begin{pmatrix}
1 & 0 \\
0 & 0 
\end{pmatrix} ,
\end{equation}
and $(\bm{\Pi}\bm{B}\bm{\Pi})^{-1}$ denotes the (Moore-Penrose) pseudoinverse. Specifically, $(\bm{\Pi}\bm{B}\bm{\Pi})^{-1} = B_{11} \bm{\Pi}$, with $B_{11}$ the top-left entry of $\bm{B}$. Hence,
\begin{Frame}[$\xh$-measurement -- post-meas. covariance]
\begin{equation}
\bs_A = \bm{A} - \frac{1}{B_{11}} \bm{C}\bm{\Pi}\bm{C}^T 
\end{equation}
\end{Frame}

The displacement vector of A after the measurement does depend on the outcome. We let $u$ denote the measurement outcome and define the vector
\begin{equation}
\bm{u} = \begin{pmatrix}
u \\ 0
\end{pmatrix} .
\end{equation}
The displacement vector of the conditional state is then
\begin{align}
\rd_A & = \bm{a} - \bm{C} (\bm{\Pi}\bm{B}\bm{\Pi})^{-1} (\bm{b} - \bm{u}) ,
\end{align}
or equivalently
\begin{Frame}[$\xh$-measurement -- post-meas. displacement]
\begin{align}
\rd_A = \bm{a} - \frac{1}{B_{11}} \bm{C}\bm{\Pi} (\bm{b} - \bm{u}) 
\end{align}
\end{Frame}

The probability distribution over outcomes only depends on the reduced state of the subsystem being measured. In the case of a single mode considered here that means it depends only on $\bm{B}$ and $\bm{b}$. For an $\xh$-measurement, the distribution is a Gaussian with variance $B_{11}$ centered at $b_1$, i.e.
\begin{Frame}[$\xh$-measurement -- outcome distribution]
\begin{equation}
\label{eq.xdist}
P_{\xh}(u) = \frac{1}{\sqrt{2\pi B_{11}}} e^{-\frac{1}{2B_{11}}(u-b_1)^2} 
\end{equation}
\end{Frame}
which is also the marginal of the Wigner function for the reduced state of B.\\

\noindent \textit{Measurements in other directions}\\

\noindent Note that, while we have given the expressions here for an $\xh$-measurement, a measurement of an arbitrary quadrature $\xh(\phi)$ can be modelled as first applying a phase shift of $-\phi$, rotating $\xh(\phi)$ to $\xh(0) = \xh$, and then measuring $\xh$. Hence, in the above expressions, one only needs to replace
\begin{align}
\bm{B} & \rightarrow \bm{R}(-\phi) \bm{B} \bm{R}^T(-\phi) , \\
\bm{C} & \rightarrow \bm{C} \bm{R}^T(-\phi) , \\
\bm{b} & \rightarrow \bm{R}(-\phi) \bm{b} ,
\end{align}
where
\begin{equation}
\bm{R}(-\phi) = \begin{pmatrix}
\cos(\phi) & -\sin(\phi) \\
\sin(\phi) & \cos(\phi) 
\end{pmatrix} .
\end{equation}
In particular, for a measurement of $\ph = \xh(\frac{\pi}{2})$ with outcome $u$, we have
\begin{equation}
\bs_A = \bm{A} - \frac{1}{\bm{B}_{22}} \bm{C}\bm{\Pi}'\bm{C}^T ,
\end{equation}
and 
\begin{equation}
\rd_A = \bm{a} - \frac{1}{\bm{B}_{22}} \bm{C} \bm{\Pi}' (\bm{b} + \bm{u}') ,
\end{equation}
where
\begin{equation}
\bm{\Pi}' = \begin{pmatrix}
0 & 0 \\
0 & 1 
\end{pmatrix} , \hspace{1cm} \bm{u}' = \begin{pmatrix}
0 \\
u 
\end{pmatrix} .
\end{equation}
The distribution over outcomes is obtained from Eq.~\eqref{eq.xdist} by replacing $B_{11} \rightarrow B_{22}$ and $b_1 \rightarrow b_2$.

\subsubsection{Other Gaussian measurements}

In addition to homodyning, other measurements may also preserve Gaussianity when acting on part of a joint Gaussian state $\rho_{AB}$. In general, for a POVM on B with elements $\{E_m\}$, the conditional state of A upon obtaining outcome $m$ may be Gaussian for some but not all $m$. It will be Gaussian whenever the characteristic function of $E_m$ is Gaussian. We list a few particular cases.\\

\noindent \textit{Projection onto the vacuum}\\

\noindent A measurement on part B of a joint Gaussian state $\rho_{AB}$ with an outcome corresponding to projection onto the vacuum state $\ket{vac}_B$ will leave system A in a Gaussian state. The covariance matrix and displacement vector of the conditional state of A are related to those of the initial state as follows.

\begin{Frame}[Vacuum projection]
For a joint system AB in a Gaussian state $\rho_{AB}$ with covariance matrix and displacement vector 
\begin{align}
\bs = \begin{pmatrix}
\bm{A} & \bm{C} \\
\bm{C}^T & \bm{B}
\end{pmatrix} , \hspace{1cm} \rd = \begin{pmatrix}
\bm{a} \\
\bm{b}
\end{pmatrix} ,
\end{align}
the conditional state of subsystem A upon projecting subsystem B onto a vacuum state $\ket{vac}_B$ is characterised by
\begin{align}
\rd_A & = \bm{a} - \bm{C}(\bm{B} + \frac{1}{2}\openone_B)^{-1} \bm{b} , \\
\bs_A & = \bm{A} - \bm{C}(\bm{B} + \frac{1}{2}\openone_B)^{-1} \bm{C}^T .
\end{align}
\end{Frame}

Since any pure Gaussian state can be obtained from the vacuum state by a suitable Gaussian unitary, projection onto such a state can be modelled by first applying the inverse unitary and then projecting onto the vacuum, using the results above. Alternatively, one can apply the expression below.\\

\noindent \textit{Projection onto a pure Gaussian state}\\

A measurement on subsystem B of a joint Gaussian state $\rho_{AB}$ with an outcome corresponding to projection onto an pure Gaussian state will leave subsystem A in a Gaussian state, which cab be obtained as follows.

\begin{Frame}[Gaussian projection]
Consider a joint system AB in a Gaussian state $\rho_{AB}$ with covariance matrix and displacement vector 
\begin{align}
\bs = \begin{pmatrix}
\bm{A} & \bm{C} \\
\bm{C}^T & \bm{B}
\end{pmatrix} , \hspace{1cm} \rd = \begin{pmatrix}
\bm{a} \\
\bm{b}
\end{pmatrix} ,
\end{align}
and consider a projection of subsystem B onto a pure Gaussian state with covariance matrix $\bm{M}$ and displacement vector $\bm{m}$. The conditional state of subsystem A is characterised by
\begin{align}
\rd_A & = \bm{a} - \bm{C}(\bm{B} + \bm{M})^{-1} (\bm{b} - \bm{m}) , \\
\bs_A & = \bm{A} - \bm{C}(\bm{B} + \bm{M})^{-1} \bm{C}^T .
\end{align}
\end{Frame}

In particular, for projections onto (products of) coherent states $\ket{\bm{\alpha}}$, we have 
\begin{align}
\bm{m} & =  \sqrt{2} \begin{pmatrix}
\Re(\bm{\alpha}) \\
\Im(\bm{\alpha}) 
\end{pmatrix} , \\
\bm{M} & =  \frac{1}{2}\openone_B .
\end{align}

\section{Gaussian integrals and fidelity}

\subsubsection{Integrals of multivariate Gaussian functions}

Here, we list some useful expressions for integrals of Gaussian functions.

\begin{Frame}[Gaussian integral - one dimension]
For $a,b,c \in \mathbb{R}$ and $a > 0$
\begin{equation}
\int_{-\infty}^{\infty} e^{-a x^2 + b x + c} dx = \sqrt{\frac{\pi}{a}}e^{\frac{b^2}{4a}+c} .
\end{equation}
\end{Frame}
In particular, for $\sigma > 0$ and $\bar{x}\in \mathbb{R}$
\begin{equation}
\int_{-\infty}^{\infty} e^{-\frac{1}{2\sigma} (x-\bar{x})^2} dx = \sqrt{2\pi\sigma} .
\end{equation}
In higher dimensions we have
\begin{Frame}[Gaussian integral - $n$ dimensions]
For $\bm{A}$ symmetric, real, positive definite and $\bm{b}\in\mathbb{R}^n$
\begin{equation}
\int_{\mathbb{R}^{n}} e^{-\frac{1}{2}\rb^T \bm{A} \rb + \bm{b}^T \rb} d\rb = \sqrt{\frac{(2\pi)^{n}}{\det(\bm{A})}} e^{\frac{1}{2}\bm{b}^T\bm{A}^{-1}\bm{b}}.
\end{equation}
\end{Frame}
In particular, with $\bs$ a covariance matrix
\begin{equation}
\int_{\mathbb{R}^{2n}} e^{-\frac{1}{2}(\rb - \rd)^T \bs^{-1} (\rb - \rd)} d\rb = \sqrt{(2\pi)^{2n}\det(\bs)} ,
\end{equation}
from which it can be seen that the Wigner function \eqref{eq.Wfctgauss} of a Gaussian state is normalised.

\subsubsection{Trace rule in phase space}

The trace of the product of two states is the integral of the product of their Wigner functions
\begin{equation}
\Tr[\rho_1 \rho_2] = (2\pi)^n \int_{\mathbb{R}^{2n}} W_1(\rb) W_2(\rb) d\rb .
\end{equation}
A similar expression holds for the characteristic functions. For Gaussian states, this leads to the following expression.
\begin{Frame}[Overlap of Gaussian states]
For $\rho_1$, $\rho_2$ Gaussian states with covariance matrices $\bs_1$, $\bs_2$ and displacement vectors $\rb_1$, $\rb_2$
\begin{equation}
\Tr[\rho_1 \rho_2] = \frac{1}{\sqrt{\det(\bm{\Sigma})}} e^{-\frac{1}{2}(\rb_1 - \rb_2)^T\bm{\Sigma}^{-1}(\rb_1 - \rb_2)} ,
\end{equation}
where $\bm{\Sigma} = \bs_1 + \bs_2$.
\end{Frame}

\subsubsection{Fidelity}

For two arbitrary mixed states $\rho_1$ and $\rho_2$ of a given system, we define the fidelity as (note that the fidelity is sometimes defined as the square of this expression)
\begin{equation}
F(\rho_1,\rho_2) = \Tr\left[\sqrt{\sqrt{\rho_1} \rho_ 2 \sqrt{\rho_1}} \right] .
\end{equation}
When both states are Gaussian, following Ref.~\citep{Banchi2015} the fidelity can be expressed in terms of their covariance matrices $\bs_1$, $\bs_2$ and displacement vectors $\rd_1$, $\rd_2$ as follows
\begin{equation}
F(\rho_1,\rho_2) = F_0(\rho_1,\rho_2) e^{-\frac{1}{4}(\rd_2-\rd_1)^T(\bs_1+\bs_2)^{-1}(\rd_2-\rd_1)}
\end{equation}
where the factor $F_0(\rho_1,\rho_2)$ (the fidelity for equal displacements) can be expressed
\begin{equation}
F_0(\rho_1,\rho_2) = \left( \frac{\det(2(\sqrt{\openone + \frac{1}{4}(\bm{V}\bm{\Omega})^{-2}} + \openone)\bm{V})}{\det(\bs_1+\bs_2)} \right)^{1/4} ,
\end{equation}
with
\begin{equation}
\bm{V} = \bm{\Omega}^T (\bs_1+\bs_2)^{-1}(\frac{1}{4}\bm{\Omega}+\bs_2\bm{\Omega}\bs_1) .
\end{equation}
Or, equivalently
\begin{equation}
F_0(\rho_1,\rho_2) = \left( \frac{\det((\sqrt{\openone - \bm{W}^{-2}} + \openone)\bm{W}i\bm{\Omega})}{\det(\bs_1+\bs_2)} \right)^{1/4} ,
\end{equation}
with
\begin{equation}
\bm{W} = -2 \bm{V} i \bm{\Omega} .
\end{equation}

In some important cases, these general expressions simplify significantly.\\

\noindent\textit{One state pure}\\

\noindent When (at least) one of the states is pure, the above expressions reduce to 
\begin{Frame}[Fidelity with pure state]
\begin{equation}
F(\rho_1,\rho_2) = \frac{1}{\sqrt[4]{\det(\bm{\Sigma})}} e^{-\frac{1}{4}(\rd_2-\rd_1)^T\bm{\Sigma}^{-1}(\rd_2-\rd_1)} ,
\end{equation}
where $\bm{\Sigma} = \bs_1 + \bs_2$.
\end{Frame}
That is, $F_0(\rho_1,\rho_2) = 1/\sqrt[4]{\det(\bs_1+\bs_2)}$ in this case. \\

\noindent\textit{Single mode}\\

\noindent When there is only a single mode (but both states may be mixed) we have
\begin{Frame}[Fidelity for single-mode mixed states]
\begin{equation}
F(\rho_1,\rho_2) = \frac{1}{\sqrt{\Gamma}} e^{-\frac{1}{4}(\rd_2-\rd_1)^T(\bs_1+\bs_2)^{-1}(\rd_2-\rd_1)} ,
\end{equation}
where
\begin{align}
\Gamma & = \sqrt{\Delta + \Lambda} - \sqrt{\Lambda} , \\
\Delta & = \det(\bs_1 + \bs_2) , \\
\Lambda & = 4(\det(\bs_1) - \frac{1}{4})(\det(\bs_2) - \frac{1}{4}) .
\end{align}
\end{Frame}

\section{Gaussian states}

In this section, we list the covariance matrices and displacement vectors for some commonly encountered Gaussian states.

We only give the expressions for a single or two modes below. They can easily be extended to multimode systems by exploiting the fact that for a product state of a composite system $\rho_{AB} = \rho_A \otimes \rho_B$, the covariance matrix is the direct sum $\bs_{AB} = \bs_A \bigoplus \bs_B$ and the displacement vector is the concatenation $\rd_{AB}^T = (\rd_A^T,\rd_B^T)$.

\subsubsection{Vacuum state}

For a single-mode system, the vacuum state $\ket{vac}$ is the ground state of the quantum harmonic oscillator corresponding, in quantum optics, to a zero-photon Fock state. For a single mode, we thus have
\begin{Frame}[Vacuum state]
\begin{align}
\bs & = \begin{pmatrix}
\frac{1}{2} & 0 \\
0 & \frac{1}{2}
\end{pmatrix} , \\
\rd & = \bm{0} .
\end{align}
\end{Frame}
For multimode systems, $\ket{vac}$ is also used to refer to the joint ground state, i.e.~a product of zero-photon states for each mode.

\subsubsection{Coherent state}

Coherent states $\ket{\alpha}$, $\alpha \in \mathbb{C}$ are displaced vacuum states, i.e.~$\ket{\alpha} = \hat{D}(\alpha)\ket{vac}$. The displacement operation preserves the variances while changing the displacement vector. Thus
\begin{Frame}[Coherent state]
\begin{align}
\bs & = \begin{pmatrix}
\frac{1}{2} & 0 \\
0 & \frac{1}{2}
\end{pmatrix} , \\
\rd & = \sqrt{2}\begin{pmatrix}
\Re(\alpha) \\
\Im(\alpha) 
\end{pmatrix} .
\end{align}
\end{Frame}

\subsubsection{Thermal state}

A canonical thermal state of a single bosonic mode with Hamiltonian $\hat{H} = \omega\ah^\dagger \ah + \frac{1}{2}$ and inverse temperature $\beta=1/T$ is given by (setting $k_B=1$)
\begin{equation}
\tau(\beta) = \frac{e^{-\beta\omega\ah^\dagger\ah}}{\Tr[e^{-\beta\omega\ah^\dagger\ah}]} .
\end{equation}
The mean number of excitations (mean number of photons) $\bar{n}$ and the temperature are related by
\begin{equation}
e^{-\beta\omega} = \frac{\bar{n}}{1+\bar{n}} .
\end{equation}
In terms of the mean photon number, in the Fock basis
\begin{equation}
\tau(\beta) = \frac{1}{1+\bar{n}} \sum_{n=0}^\infty \left(\frac{\bar{n}}{1+\bar{n}}\right)^n \ket{n}\bra{n} .
\end{equation}
The covariance matrix and displacement vector are
\begin{Frame}[Thermal state]
\begin{align}
\bs & = \begin{pmatrix}
\bar{n} + \frac{1}{2} & 0 \\
0 & \bar{n} + \frac{1}{2}
\end{pmatrix} , \\
\rd & = \bm{0} .
\end{align}
\end{Frame}
Note that the zero-temperature thermal state is the vacuum state.

\subsubsection{Single-mode squeezed vacuum state}

Single-mode squeezed vacuum is obtained by applying the single-mode squeezing operator to the vacuum, i.e.~$\ket{S(\xi)} = \hat{S}(\xi)\ket{vac}$. Using the symplectic transformation for single-mode squeezing with $\xi = r e^{i\theta}$,
\begin{Frame}[Single-mode squeezed vacuum]
\begin{align}
\bs & = \frac{1}{2} \cosh(2r) \openone_2 - \frac{1}{2}\sinh(2r)\bm{S}_\theta , \\
\rd & = \bm{0} ,
\end{align}
where
\begin{equation}
\bm{S}_\theta = \begin{pmatrix}
\cos(\theta) & \sin(\theta) \\
\sin(\theta) & -\cos(\theta)
\end{pmatrix} .
\end{equation}
\end{Frame}
In particular, for squeezing along $\xh$ ($\theta = 0$) we get
\begin{equation}
\bs = \begin{pmatrix}
\frac{1}{2} e^{-2r} & 0\\
0 & \frac{1}{2} e^{2r}
\end{pmatrix} ,
\end{equation}
and for squeezing along $\ph$ ($\theta = \pi$) we have
\begin{equation}
\bs = \begin{pmatrix}
\frac{1}{2} e^{2r} & 0\\
0 & \frac{1}{2} e^{-2r}
\end{pmatrix} .
\end{equation}

\subsubsection{Two-mode squeezed vacuum state}

Two-mode squeezed vacuum is obtained by applying the two-mode squeezing operator to the joint vacuum state of a two-mode system, i.e.~$\ket{S_2(\xi)} = \hat{S}_2(\xi)\ket{vac}$ with $\xi = r e^{i\theta}$. Using the symplectic transformation for two-mode squeezing one finds
\begin{Frame}[Two-mode squeezed vacuum]
\begin{align}
\bs & = \begin{pmatrix}
\frac{1}{2} \cosh(2r) \openone_2 & -\frac{1}{2}\sinh(2r)\bm{S}_\theta \\
-\frac{1}{2}\sinh(2r)\bm{S}_\theta & \frac{1}{2} \cosh(2r) \openone_2
\end{pmatrix} , \\
\rd & = \bm{0} ,
\end{align}
where
\begin{equation}
\bm{S}_\theta = \begin{pmatrix}
\cos(\theta) & \sin(\theta) \\
\sin(\theta) & -\cos(\theta)
\end{pmatrix} .
\end{equation}
\end{Frame}
In particular, for $\theta=0$ we have
\begin{equation}
\bm{S}_\theta = \begin{pmatrix}
1 & 0 \\
0 & -1
\end{pmatrix} ,
\end{equation}
and $\text{Var}(\xh_1+\xh_2) = \text{Var}(\ph_1-\ph_2) = e^{-2r}$.

\section{Numerical tools}

When working with large systems or complex entangled states, numerical methods may be required.

\begin{itemize}
\item QuGIT: A Python numerical toolbox for  simulating multimode Gaussian states and operations \cite{qugit2022-arxiv}.
\end{itemize}

\section{Comments and suggestions}

Any comments and corrections are welcome, as are suggestions for further relevant results and expressions which might be included, or additional numerical toolss.

\bibliography{gaussian_states_and_ops}

\end{document}